\documentclass[sigconf]{acmart}
\usepackage{makecell}
\usepackage{multirow}
\usepackage{balance}

\newcommand\blfootnote[1]{%
  \begingroup
  \renewcommand\thefootnote{}\footnote{#1}%
  \addtocounter{footnote}{-1}%
  \endgroup
}

\AtBeginDocument{%
  \providecommand\BibTeX{{%
    \normalfont B\kern-0.5em{\scshape i\kern-0.25em b}\kern-0.8em\TeX}}}

\setcopyright{acmlicensed}
\copyrightyear{2023} 
\acmYear{2023} 
\acmDOI{10.1145/3624918.3625328}

\acmConference[SIGIR-AP '23]{Annual International
ACM SIGIR Conference on Research and Development in Information Retrieval in
the Asia Pacific Region}{November 26--28, 2023}{Beijing, China}
\acmBooktitle{Annual International ACM SIGIR Conference on Research and
Development in Information Retrieval in the Asia Pacific Region (SIGIR-AP '23),
November 26--28, 2023, Beijing, China}
\acmPrice{15.00}
\acmISBN{979-8-4007-0408-6/23/11}

\begin{document}

\title{Boosting legal case retrieval by query content selection with large language models}

\author{Youchao Zhou, Heyan Huang\footnotemark, Zhijing Wu}

\affiliation{
 School of Computer Science and Technology, Beijing Institute of Technology \\ 
Southeast Academy of Information Technology, Beijing Institute of Technology\\
Beijing Engineering Research Center of High Volume Language Information Processing and Cloud Computing Applications
  \city{}
   \country{}
 }
 \email{{yczhou,hhy63,zhijingwu}@bit.edu.cn}

\renewcommand{\shortauthors}{Zhou, et al.}

\begin{abstract}
Legal case retrieval, which aims to retrieve relevant cases to a given query case, benefits judgment justice and attracts increasing attention. Unlike generic retrieval queries, legal case queries are typically long and the definition of relevance is closely related to legal-specific elements. Therefore, legal case queries may suffer from noise and sparsity of salient content, which hinders retrieval models from perceiving correct information in a query.  While previous studies have paid attention to improving retrieval models and understanding relevance judgments, we focus on enhancing legal case retrieval by utilizing the salient content in legal case queries. We first annotate the salient content in queries manually and investigate how sparse and dense retrieval models attend to those content. Then we experiment with various query content selection methods utilizing large language models (LLMs) to extract or summarize salient content and incorporate it into the retrieval models. Experimental results show that reformulating long queries using LLMs improves the performance of both sparse and dense models in legal case retrieval.

\end{abstract}






\maketitle

\blfootnote{* Corresponding Author}
\section{Introduction}

Legal case retrieval plays a critical role in intelligent legal systems, which improves the efficiency of legal professionals, for example, helping judges and lawyers to get similar legal cases. 
% To promote judgement fairness and justice, similar cases should be given similar judgments. 
Researchers have built several benchmarks (e.g., LeCaRD~\cite{lecard} in Chinese and COLIEE~\cite{COLIEE} in English) and hold competitions~\cite{cail2019,COLIEE} to promote research on legal case retrieval.

\begin{figure}[t]
  \centering
  \includegraphics[width=\linewidth]{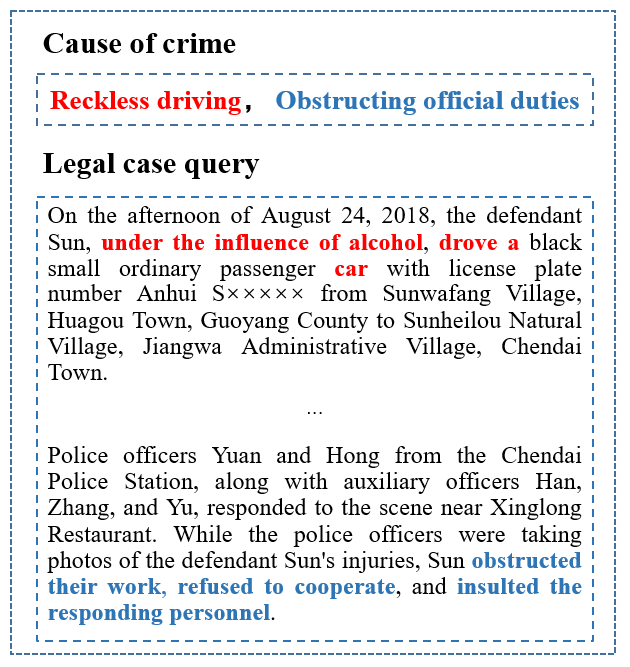}
  \setlength{\abovecaptionskip}{-0.5em}
  \caption{An example of a legal case query (translated from Chinese) in LeCaRD. The highlighted content represents the four criminal elements which matches the cause of crime in the same color. They are potentially more valuable for legal case retrieval since they have decisive impact on legal judgment.}
  \vspace{-1em}
  \label{fig: query_example}
\end{figure}

Legal case retrieval is more complex and domain-specific in comparison to generic ad hoc retrieval tasks ~\cite{legalrj}. The definition of case relevance requires legal knowledge, and the documents are extremely lengthy.

However, not all passages in a query contribute the same to document relevance judgment ~\cite{passagerole}. Figure ~\ref{fig: query_example} shows an example of a query in LeCaRD. The highlighted content (see section ~\ref{sec:dataset} for more details) represents the four criminal elements, which have a decisive impact on the case judgment and thus they are potentially more valuable for legal case retrieval. As shown in Figure~\ref{fig: query_example}, a long query contains redundant text, leading to the sparsity of salient content. Existing dense retrieval models ~\cite{bertpli,salier,legalFE} 
may suffer from the sparsity of salient content and fail to correctly perceive the valuable information. According to \citet{2023thuir}, dense models sometimes exhibit worse performance compared to traditional sparse retrieval models.

Existing research in legal case retrieval primarily focuses on improving retrieval performance from the model perspective ~\cite{bertpli,salier} and user perspective ~\cite{legalrj}. Several studies explore the effectiveness of query summarization in improving retrieval performance. For example,
\citet{LCsum} train a phrase scoring model to extract key sentences to construct an embedding as the summary of a query. \citet{2022LeiBi} leverage unsupervised methods to extract keywords and train a supervised longformer-based model to summarize a query. However, those methods do not stably improve the performances of dense retrieval models and they do not provide a further understanding of legal case queries. 

The ability of models to capture valuable information in the presence of redundant textual information within legal case queries remains to be investigated. In this paper, we refer to valuable textual information within legal case queries as \textbf{salient content}, and we propose the first research question:

\begin{itemize}
\item {\bfseries RQ1}: How do traditional sparse retrieval models and dense retrieval models attend to salient content in a legal case query?
\end{itemize}

To address RQ1, we first recruit lawyers to annotate the salient content of queries in LeCaRD dataset. 
% \begin{itemize}
% \item Notice that we also refer to salient content as annotated words in the following part.
% \end{itemize}
Then we implement classical sparse and dense retrieval models for further comparison to understand model differences from word frequency and entropy views. We find both models can perceive salient content but they are not correlated well with humans. This motivates our second research question:

\begin{itemize}
\item {\bfseries RQ2}: Can we reformulate a legal case query to highlight salient content and improve retrieval models?
\end{itemize}

In this paper, we mainly reformulate queries through \textbf{query content selection} to highlight salient content. We utilize large language models(LLMs) to extract or summarize salient content and incorporate it into the retrieval models. Experimental results show that highlighting salient content in legal case queries can boost retrieval models. To summarize, the main contributions are as follows:

\begin{itemize}
% \item  We show sparse and dense retrieval models perceive salient content in a query in quite dissimilar views.

% \item Experimental results show the effectiveness of performing query content with LLMs, which highlight salient content and improve both sparse and dense models in legal case retrieval.\footnote{Annotation data and codes will be released after review.}

\item We collect annotations for salient content based on LeCaRD dataset (Section~\ref{sec:dataset}) and conduct an in-depth analysis of how existing sparse and dense models perceive the salient content (Section~\ref{sec:analysis}).
\item We experiment with various query content selection methods to obtain the salient content within queries and utilize it to effectively improve the performance of retrieval models (Section~\ref{sec:retrieval}). \footnote{Annotations and analysis codes are available at 

https://github.com/zuochao912/LegalsearchSum}

\end{itemize}

\section{Related work}
\subsection{Legal case retrieval}
The research on legal case retrieval has a long history, and previous methods primarily focused on manually creating features for ranking models to effectively leverage expert knowledge ~\cite{hislegalir}. With the rise of pre-trained models, researchers have become more interested in NLP methods. Generally, dense retrieval models are popular as they perform semantic matching. They need to be trained in supervised settings \cite{bertmtft,legalFE}. However, since relevance labels require legal experts manually annotating, the scale of dataset is limited and thus contains limited legal relevance knowledge. To better leverage legal documents, researchers perform further pretraining to enhance the semantics of text embedding ~\cite{bertpli,salier}. 

While the methods mentioned above rely on top-down supervised relevance signals or bottom-up structure signals to enhance models, some researchers have noticed the characteristic of lengthy legal texts. They extract key content or summarize legal docs to perform retrieval experiments ~\cite {LCsum,Lformersum,legalqg}. However, those methods mainly benefit lexical-based models, and further exploration for dense models is needed. Besides,\citet{exLegalIR}  construct an explainable legal case matching dataset and design models which can also extract rationales from paired legal cases. However, they do not investigate how retrieval models perceive query differs. Thus we provide a further understanding of the model perception.
 
\subsection{Query reformulation and content selection}
Query reformulation is mainly categorized into query expansion and query rewriting, as some queries appear to be vague and noisy to express user intents ~\cite{querynoise}. The former enhances the information retrieval effectiveness and the latter solves the term mismatch problem mainly caused by polysemy and synonymy ~\cite{CoopQrir}. Meanwhile, a few researchers investigate how to extract valuable content in query for the effectiveness of retrieval model. \citet{Disckw} identify the key concepts in verbose keyword queries to improve lexical based models. However, legal queries are quite dissimilar as they are typically long. Models may suffer from the sparsity of salient content especially when relevance label provide limited guidance. 

Performing salient content selection is a nature way to alleviate the problem and it has been widely used in other fields ~\cite{longdocsum}. For example, \citet{BUsum} apply extract-abstract methods to improve summary performance. They utilize a pre-trained pointer-generator content selector to mask not salient content, which makes summary model simpler to train. \citet{BUTD} leverages a pre-trained object detector to select salient image regions for visual querstion answer(VQA) model and achieve championship in 2017 VQA competition.

Recently, LLMs are used to perform query reformulation as they possess a vast amount of world knowledge. For example, \citet{hyde} leverage LLMs to expand queries to improve unsupervised dense retriever. \citet{LLM4convsearch} explore query rewriting with simple prompts of GPT-3 to represent the user’s real contextual search intent in conversational search, which achieve significant improvement. Considering the success of LLMs, we mainly explore reformulating queries by content selection with LLMs to boost retrieval models. 

\section{preliminaries}
In this section, we first describe how we annotate salient content of legal case queries. Then we implement prevalent models in legal case retrieval for further analysis.  

\subsection{Dataset}
\label{sec:dataset}

We collect annotations based on the LeCaRD dataset ~\cite{lecard}, a well-annotated Chinese legal case retrieval benchmark. LeCaRD composes 107 queries and 10,700 candidate cases selected from a corpus of over 43,000 Chinese criminal judgments. LeCard covers queries across diverse difficulties and categories and the top 30 relevant candidate cases of each query case are annotated in four-level relevance scales. Though the relevance is annotated according to key circumstances and key elements, the dataset does not provide the relevance rationale.

We recruit two lawyers from a law office to annotate the salient content in queries. They are senior lawyers with over 20 years of experience in the field and are familiar with both criminal and civil laws. They are capable of handling annotation tasks.

In Statute Law System of China, for example, nearly all criminal judges make decisions based on the Four Elements Theory ~\cite{legalqa}, including the subject, the object, the conduct, and the mental state. Since those elements are naturally more valuable in a legal case, we ask annotators to mainly focus on the four elements in queries. 

Specifically, the annotation procedures are as follows. When annotating, we show a query and the corresponding crime charges at the same time. We ask annotators to firstly focus on the four criminal elements and then rethink whether they are helpful for legal case retrieval. For example, the subject which refers to the person or organization who has committed the criminal offense does not have reference value for legal case retrieval and may not be annotated as salient content. We ask them to keep annotations that suit the crime best because annotators report that sometimes there is an excessive amount of annotation in a query. Besides, we ask them to maintain complete linguistic units, which means the annotation units should be phrases or short sentences. They discuss to reach a consensus and finally complete the annotation.

The statistics of annotations are shown in Table \ref{tab:annot_sta}. We calculate the compression rate, which is the length ratio of salient content and original query. The average compression rate is relatively low, which means the most informative content for legal judgement and potentially valuable for retrieval task in a query is very limited. 

\begin{table}[t]
  \centering
  \caption{Statistics of query annotations.}
    \begin{tabular}{lll}
    \toprule
    Statistic & Have stop words & Value \\
    \midrule
    Avg. query  & Yes   & 495.5 \\
    length      & No    & 444.6 \\
    \midrule
    Avg. annotation  & Yes   & 72.14 \\
    length      & No    & 60.77 \\
    \midrule
    Avg. query  & Yes   & 19.06 \\
    compression rate(\%)      & No    & 14.52 \\
    \bottomrule
    \end{tabular}%
  \label{tab:annot_sta}%
\end{table}%

\subsection{Comparison models}

\subsubsection{Model introduction}
We implement two types of retrieval models widely used in legal case retrieval. Namely, the traditional sparse retrieval models based on lexical matching and dense retrieval models based on semantic matching of PLMs. 
Dense retrieval models show better performance in most cases, while the traditional models still act as a strong baseline in legal search ~\cite{2023thuir}. We give a brief introduction as follows.

\begin{itemize}
\item Sparse retrieval models
\begin{itemize}
    \item  TF-IDF ~\cite{tfidf}: A bag-of-words model using the TF-IDF value of each word to construct the document representation vector.
    \item  BM25 ~\cite{bm25}: An effective exact word matching retriever based on TF and IDF value.
    \item  QL ~\cite{statIR}: An n-gram language model retriever based on Dirichlet smoothing, which regards the generation probability of a query by a document as similarity metrics.
    
\end{itemize}

\item Dense retrieval models
\begin{itemize}
    \item BERT-CLS ~\cite{bertcls}: A simple BERT cross encoder for ranking, which includes a Fully Connect Feedforward Network(FFN) over a BERT encoder. As the FFN leverages representation of [CLS] to learn, we refer to it as BERT-CLS.
    \item BERT-PLI ~\cite{bertpli}: A complex cross encoder that conducts passage-level interaction based on paragraph sentence embedding, which won the second place in a legal search competition.
\end{itemize}
\end{itemize}

For dense retrieval models, we use BERT as the backbone because it is still prevalent in finetuning or pretraining process. Besides, we adopt interaction-based cross encoders for dense retrieval models as they generally perform much better than representation-based models when they are only fine-tuned on the datasets. 

\begin{table*}[htbp]
  \centering
  \caption{Evaluation of comparison models on LeCaRD. The best performance of each model type for each metric is marked in bold. The number in the bracket after a model name denotes the input format of the model. }
    \begin{tabular}{clllllll}
    \toprule
    \multicolumn{1}{c}{Model Type} & Model & \multicolumn{1}{l}{P@5(\%)} & \multicolumn{1}{l}{P@10(\%)} & \multicolumn{1}{l}{MAP(\%)} & \multicolumn{1}{l}{NDCG@10(\%)} & \multicolumn{1}{l}{NDCG@20(\%)} & \multicolumn{1}{l}{NDCG@30(\%)} \\
    \midrule
    \multicolumn{1}{c}{\multirow{3}[2]{*}{\makecell{Traditional sparse \\ retrieval model}}} & TF-IDF & 33.08 & 31.78 & 41.76 & 64.95 & 70.57 & 78.13 \\
          & QL    & \textbf{42.24} & \textbf{40.37} & \textbf{48.84} & \textbf{74.27} & \textbf{79.08} & \textbf{87.19} \\
          & BM25  & 40.56 & 37.85 & 47.8  & 72.05 & 77.89 & 86.51 \\
    \midrule
    \multicolumn{1}{c}{\multirow{3}[2]{*}{\makecell{Dense retrieval \\ model}}} & BERT-CLS(256,256) & \textbf{46.16} & \textbf{41.87} & \textbf{55.41} & 77.6  & 80.6  & 85.19 \\
          & BERT-PLI(1*200,3*100) & 43.18 & 40    & 51.27 & 77.41 & 80.53 & 85.42 \\
          & BERT-PLI(1*200,12*100) & 43.36 & 39.9  & 52.55 & \textbf{77.75} & \textbf{81.63} & \textbf{86.03} \\
    \bottomrule
    \end{tabular}%
  \label{tab:origin_perform}%
\end{table*}%

\subsubsection{Set up}
\label{sec:setup_original}
As LeCaRD dataset does not provide an official train-test split, We perform a 5-fold cross-evaluation and run various random seeds for dense retrieval models. 

We implement the sparse retrieval models following \citet{lecard}. We tune the BM25 model and set the parameters $k=1.4$ and $b=0.6$. We tune the QL model and set the interpolation parameters $\lambda=0.1$.

For dense retrieval models, we use bert-base-chinese\footnote{https://huggingface.co/bert-base-chinese} as the backbone. For BERT-CLS, we build a simple fully connected network over BERT encoder, which contains 2 hidden layers. For BERT-PLI, We do not pre-train the model as \citet{bertpli}, we only adopt its architecture and directly finetune it on the dataset.

As queries and documents are generally longer than the input of BERT-CLS, we truncate the inputs and the document. For BERT-CLS, the query and document are both truncated to 256 character lengths following \citet{lecard}. For BERT-PLI, we truncate the query to 200 characters and the document to 100 characters for each text segment. The inputs of dense models are ordered as $[CLS] \, q \, [SEP] \, d \, [SEP]$, where $q$ represents query tokens and d for document tokens.

For those models, we take the representation of [CLS] token as inputs of modules over the BERT encoder. We just follow the \citet{bertpli,lecard}, training the model with cross entropy in a pointwise style. We use an Adam optimizer with linear warmup learning rate organization methods to finetune each model in 6 epochs. We set warmup ratio to 0.1 and the learning rate to 3e-5.

\subsubsection{Original model performance}

Following \citet{lecard}, we adopt the same evaluation metrics, including precision metrics and ranking metrics. All the results are shown in Table \ref{tab:origin_perform}, which reflects the same trend as the original paper ~\cite{lecard}. The number in the bracket after a model name denotes the input format. For example, BERT-PLI(1*100,12*100) denotes that the input for BERT-PLI includes 1 segment of query whose length is 100 and 12 segments of documents whose length is 100.

Traditional sparse retrieval models generally perform not as well as dense retrieval models as they are unsupervised. TF-IDF performs much worse as it is a bag-of-words model and only contains the word frequency information and does capture many matching signals. BM25 and QL perform better as they not only perceive word frequency but also perceive matching signals between a query and a document.

Dense retrieval models perform well on both precision and ranking metrics. Though BERT-CLS is a simple model, it performs significantly better than other models especially on the precision metrics. We find that when the input has a similar length, BERT-PLI does not perform as well as BERT-CLS. This could be the result of  the separation of a document, which leads to the loss of interaction between passages. Despite we increase the inputs documents passages for BERT-PLI, we do not observe significant improvement. That may be the limited relevance labels provide insufficient guiding signals when BERT-PLI faces excessive textual inputs. This explains why truncating inputs of legal query is still prevalent in competitions. 

\begin{figure}[t]
  \centering
  \includegraphics[width=\linewidth]{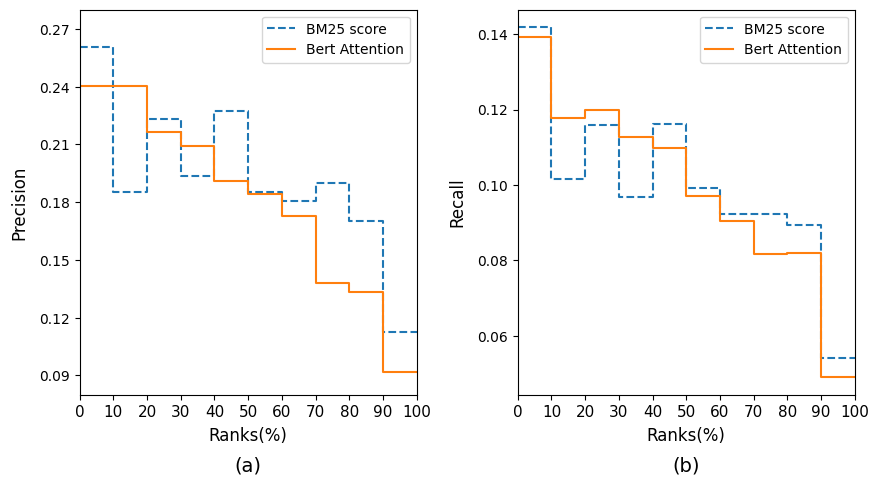}
  \caption{(a) precision and (b) recall of salient words on ranking intervals of query words according to the model attention.}
  \Description{precision and recall rate}
  \label{fig: attention_metric}
\end{figure}

\section{Retrieval model analysis}
\label{sec:analysis}

In this section, we answer RQ1: How do traditional sparse retrieval models and dense retrieval models attend to salient content in a legal case query? We select representative models and analyze them based on the annotated salient content within a query.  

\subsection{Model attention Definition}
We define model attention as the indicator of content importance from the perspective of models. We choose BM25 and BERT-CLS as representatives of each type of model for analysis as they perform well and are suitable to calculate model attention. Notice we do not choose QL as the importance of query word is calculated to a candidate document. 

For BM25, model attention is defined as word importance of BM25 score used in matching progress. Here we calculate the BM25 score for each word in a query following \citet{legalrj}. Given a query $q$ containing $n$ words $q=\{s_1,s_2,...,s_n \}$, the importance of a word $s_j$ is calculated using the BM25 matching score:

\begin{equation}
  \omega(s_j,q)= IDF(s_j) \cdot \frac{k+1}{TF(s_j,q)+k\cdot(1-b+b\cdot\frac{|S|}{avgl})},
\end{equation}
where  $IDF(s_j)$ is calculated on 46,000 documents from LeCaRD corpus, $TF(s_j,q)$ is the frequency of $s_j$ in query $q$, $|s|$ is the length of  $q$ and $avgl$ is the average length of all queries.

For BERT-CLS, model attention is defined as BERT attention, which generally used to explain the perception of a model. BERT attention is complicated as BERT uses multi-head attention for each layer. As the attention of last layer has more semantic meaning ~\cite{Bertlook}, and the attention of each token tend to allocate more value to the same token ~\cite{HowBertrank} in the texts. Following \citet{Bertlook}, we use the average attention of heads in the last layer from [CLS] token as the model attention. We denote the attention weight from $[CLS]$ token to other query token $t_j$ as $w_j$.  As the query length varies, we perform min-max normalization to get a normalized attention weight $w_j'$. Finally We align the $w_j'$ to each query word $s_i$ to get word attention according to:
\begin{equation}
    atten(s_i)=\frac{1}{len(s_i)} \sum_j|s_i \cap t_j| \cdot {\frac{w_j'}{len(t_j)}}, 
\end{equation}
where $s_i \cap t_j$ denotes the intersection character set between token $t_j$ and word $s_i$. As the attention distribution is influenced by the relevance between a query and a document, we only choose the most relevant documents of label 3 to evaluate the $[CLS]$ token attention toward query words. 

\begin{figure}[t]
  \centering
  \includegraphics[width=\linewidth]{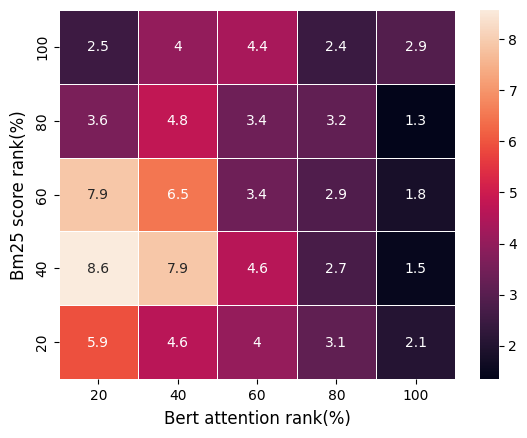}
  \caption{Joint distributions of the salient words' ranks according to BM25 score and BERT-CLS attention. The numbers in the cells represent the proportions(\%) of salient words in the corresponding ranking intervals.}
  \label{fig: attention_diff}
\end{figure}

\subsection{Model attention analysis}

We regard the salient content annotated by lawyers (see Section~\ref{sec:dataset} for details) as ground truth. Words in salient content are denoted as \textbf{salient words}.
We rank query words according to the model's attention and calculate the precision and recall of salient words to evaluate the perception performance of a model. As the query lengths vary, we calculate the precision and recall on intervals of word rankings. The results are shown in Figure ~\ref{fig: attention_metric}.

According to the precision and recall curve of models, we observe that both BM25 score and BERT attention indeed can be regarded as the model attention to queries since they place annotated salient words in the front and less valuable words in the back.

However, both models do not correlate with human annotation well as they do not achieve good precision performance. Despite their best performance in the top 10\verb|%| ranking interval, the precision of BM25 score is only 26\verb|%|, slightly higher than 19\verb|%|,which is the average compression rate of salient content in a query.

\begin{figure}[t]
  \centering
  \includegraphics[width=\linewidth]{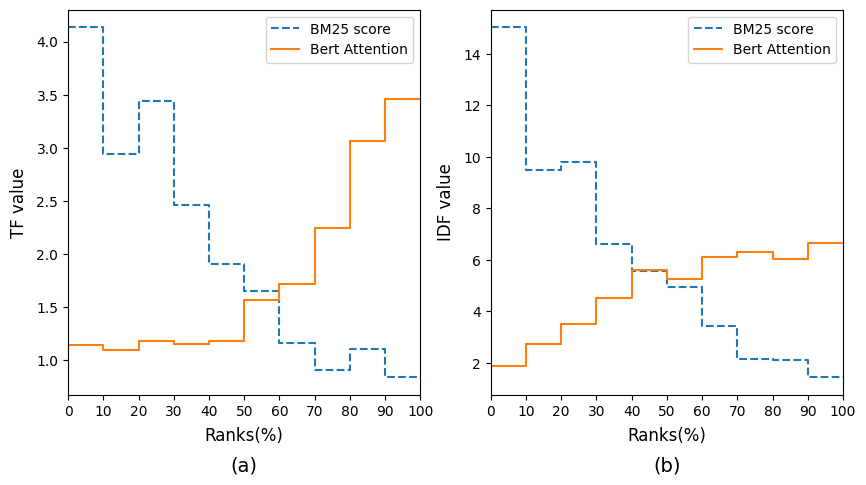}
  \caption{Average (a) TF and (b) IDF value of salient words in the ranking intervals according to the model attention.}
  \Description{A woman and a girl in white dresses sit in an open car.}
  \label{fig: attention_detail}
\end{figure}

Though the performances of model attention appears close to each other, the matching mechanism is actually distinct. We compare the ranking discrepancy for the same annotation by two dissimilar models. Notice that the perception fields of those two models are distinct. For BM25, it perceives the whole query while BERT-CLS only perceives the truncated part. Therefore, we only analyze the overlapping truncated part here. The results are shown in Figure \ref{fig: attention_diff}. The numbers in the cells represent the proportions(\%) of salient words in the corresponding ranking intervals of the two models. Models perceive annotation in separate ways as the values on the main diagonal are not higher than some non-diagonal cells. In addition, BERT attention are more sensitive in the truncated part as the elements of the upper triangular matrix have larger numerical values than the elements of the lower triangular matrix. However, according to the recall curves in \ref{fig: attention_metric}, they show a similar value, indicating Bert omit the salient content outside the truncated part.

\begin{table*}[t]
  \centering
  \caption{Evaluation of comparison models on different types of query. The best result of each model is marked in bold. The best results among all models and query types except for annotation are highlighted with underlines. Values in brackets represent increments of metrics compared to the original query.}
    \begin{tabular}{clllllll}
    \toprule
    Model & Query type & P@5(\%) & P@10(\%) & MAP(\%) & NDCG@10(\%) & NDCG@20(\%) & NDCG@30(\%) \\
    \midrule
    \multirow{5}[2]{*}{BM25} & original & 40.56 & 37.85 & 47.8  & 72.05 & 77.89 & 86.51 \\
          & keyword & \textbf{42.99(2.43)} & 39.44(1.59) & 48.8(1) & 73.84(1.79) & 78.86(0.97) & 84.93(-1.58) \\
          & key sentence & 42.62(2.06) & 40(2.15) & \textbf{50.78(2.98)} & 74.58(2.53) & 79.44(1.55) & \textbf{86.61(0.1)} \\
          & summary & 42.24(1.68) & \textbf{40.75(2.9)} & 50.13(2.33) & \textbf{75.82(3.77)} & \textbf{80.11(2.22)} & 86.19(-0.32) \\
          & annotation & 44.49 & 41.31 & 52.52 & 77.05 & 81.64 & 86.56 \\
    \midrule
    \midrule
    \multirow{5}[2]{*}{BERT-CLS} & original & 46.16 & 41.87 & 55.41 & 77.6  & 80.6  & 85.19 \\
          & keyword & 47.29(1.13) & 41.21(-0.66) & \textbf{56.81(1.4)} & 79.29(1.69) & 82.32(1.72) & 86.65(1.46) \\
          & key sentence & 46.54(0.38) & 40.93(-0.94) & 54.97(-0.44) & 79.63(2.03) & 82.89(2.29) & \textbf{87.06(1.87)} \\
          & summary & \textbf{\underline{48.97(2.81)}} & \textbf{42.71(0.84)} & 56.64(1.23) & \textbf{80.22(2.62)} & \textbf{82.96(2.36)} & 86.76(1.57) \\
          & annotation & 49.53 & 44.02 & 58.8  & 80.83 & 82.99 & 86.54 \\
    \midrule
    \midrule
    \multirow{5}[2]{*}{BERT-PLI} & original & 43.36 & 39.9  & 52.55 & 77.75 & 81.63 & 86.03 \\
          & keyword & 46.54(3.18) & 41.86(1.96) & 55.34(2.79) & 80.1(2.35) & 82.91(1.28) & 87.25(1.22) \\
          & key sentence & 44.3(0.94) & 41.4(1.5) & 53.92(1.37) & 80.18(2.43) & 83.52(1.89) & 88.37(2.34) \\
          & summary & \textbf{48.03(4.67)} & \textbf{\underline{43.64(3.74)}} & \textbf{\underline{58.24(5.69)}} & \textbf{\underline{81.33(3.58)}} & \textbf{\underline{84.21(2.58)}} & \textbf{\underline{88.81(2.78)}} \\
          & annotation & 50.02 & 44.55 & 59.8  & 81.39 & 84.45 & 88.65 \\
    \bottomrule
    \end{tabular}%
  \label{tab:main_results}%
\end{table*}%

To further understand the differences, we analyze the model attention from the perspective of word frequency and entropy, i.e. term frequency(TF) and inverted document frequency(IDF) value of words. We calculate the average TF and IDF value of words on each ranking interval. As shown in Figure \ref{fig: attention_detail}, we notice that models are sensitive to different parts of salient words. BM25 scores prefer words with high TF and IDF values while BERT attention prefer words with low TF and IDF values. This phenomenon means BERT attentions prefers words rare in queries and common among documents while BM25 is exactly the opposite. A similar phenomenon is observerd in \cite{TransScan}, where self-attention shows Discriminative Bias, which will pays attention to distinct tokens that appear uniquely in token producing process, while loses interest to common tokens that appear across multiple next tokens.

\section{legal case retrieval based on query reformulation}
\label{sec:retrieval}

In this section, we answer RQ2: Can we reformulate a legal case query to highlight salient content and improve retrieval models? Previous research usually train a supervised summary model ~\cite{bertpli,LCsum} or extract keywords using unsupervised methods like KLI and TextRank \cite{2022LeiBi}. However, they often fail to improve retrieval models. Considering the limited knowledge of those methods and complexity of salient content in query, we leverage LLMs to reformulate queries. Notice we mainly reformulate queries through query content selection.

\subsection{Query reformulation methods}

We mainly design three kinds of methods to provide candidate salient content, namely keyword extraction, key sentence extraction, and summary. Specifically, we use GPT-3.5-turbo ~\cite{gpt3.5} to generate those reformulated queries. We design corresponding prompts for each task according to structured prompts\footnote{https://www.promptingguide.ai/zh} techniques. Those prompts mainly consist of four parts, namely role player, task explanation, task requirements, and details. We design very simple prompts to reformulate queries in a zero-shot setting. We set all the role players with the prompt text: "You are a legal expert with knowledge of the Chinese law”.  The task explanation is set according to the task. We set the task requirements mainly to control the format of generation. As we find that GPT-3.5-turbo is not aware of the legal judgment concept words such as "key elements" , we add a description to that concept, i.e. "Pay attention to the key statement that plays a crucial role in the case judgment."  The prompts are shown as follows.

\begin{itemize}
\item {\bfseries Keyword extraction}: You are a legal expert with knowledge of the law. You need to do the keyword extraction task of the law for keyword extraction. Pay attention to the key statement that plays a crucial role in the case judgement. Please separate each word using comma.

\item {\bfseries Key sentence extraction}:You are a legal expert with knowledge of the law. You need to do the legal key content extraction task for key sentence extraction. Pay attention to the key statement that plays a crucial role in the case judgement. Please list the key sentence.

\item {\bfseries Summary}:You are a legal expert with knowledge of the law. You need to make a summary of the above legal documents. Pay attention to the key statement that plays a crucial role in the case judgement.

\end{itemize}

\subsection{Experimental settings}

We adopt representative models including BM25, BERT-CLS(256, 256) for ensuring consistency as Section ~\ref{sec:analysis}. We also adopt BERT-PLI
(1*200,12*100) as it does not perform well on original queries in a purely finetuning setting. The model configuration follows the parameters configuration in Section ~\ref{sec:setup_original}, and the only difference is the input query for a fair comparison to the original queries. 

We use the default parameters of the OpenAI chat-completions api\footnote{https://platform.openai.com/docs/guides/gpt/chat-completions-api} to generate the query with the prompts designed above and conduct experiments on those reformulated queries with representative retrieval models. For dense retrieval methods, as keywords and key sentences are not fluent nature language, thus we reorder the inputs. For extracted keyword queries, we order inputs as follows:
$$[CLS]\, Keywords:\, s_1\, ,...,s_n\, [SEP]\, doc\, [SEP],$$where $s_i$ is the $i^{th}$ word in reformulated query.

For extracted key sentences, we use the minimum edit distance to match the original sentence and make sure the sentence order is the same as the original text. Following \citet{uqa}, we separate the sentence using a separate sign \verb|\n| and order the inputs as follows:
$$[CLS] S_1 \, \verb|\n| \, S_2 \, \verb|\n| \, ,..., \, \verb|\n| \, S_k \, [SEP] \, doc \, [SEP],$$ where $S_j$ is the $j^{th}$ sentence in reformulated query.

Except for queries generated by LLMs, we also perform retrieval models on the annotated salient content as query to verify the effectiveness. The extracted annotations are not fluent and contain different granularities of linguistic units such as words, phrases, and short sentences. Here we try to uniform those linguistic units by choosing the short sentence that contains annotation and connecting them with proper punctuation.

\subsection{Experimental results}

Table ~\ref{tab:main_results} displays overall results on LeCaRD in five-fold cross validation settings. We conduct each experiment with 5 random seeds and report an average performance. 

BM25 model stably improves over reformulated queries, indicating that even in a zero-shot setting, LLMs can highlight salient words to promote the retrieval models. Dense retrieval models also improve on keywords and key sentence queries, indicating that with a moderate ordering format of discontinuous linguistic units, dense models also perform word matching and ranking.

From the perspective of retrieval models, query content selection methods have various influences over retrieval models. For BM25, the key sentence query improves precision metrics most while some queries fail on ranking metrics. As BM25 utilize word level information without semantics information, the performance varies on different metrics. BERT-CLS model performs the best on the summarized query on nearly all metrics. While on other reformulated queries, BERT-CLS even drops on some precision metrics, indicating the reformulated queries do not stably improve all models as it may still miss potentially salient content. For BERT-PLI model, all kinds of reformulated query help improve on all metrics and summarized query improves the precision metrics significantly. Compared to the moderate increments of BERT-CLS on summary query,  the sharp increments of BERT-PLI indicate that queries with appropriate information density help the complicated model learn.

From the perspective of query reformulation methods, we find nearly all retrieval models are improved. The summarized queries stably improve all retrieval models and especially for BM25 and BERT-PLI. This might be the summary generated by LLMs preserve valuable information and in fluent nature language format. The annotated salient content performs the best especially in precision-based metrics, despite its low compression rate according to Table ~\ref{tab:annot_sta}. The performance of models on annotation demonstrate the effectiveness of properly selecting salient content, indicating that salient part of original query can satisfy relevance judgement.

\subsection{Discussion}

To further explain the relationship between query type and model performance, we evaluate the recall of salient content with the degree of similarity between those reformulated queries and the human-annotated salient words. Specifically, we calculate the overlap between annotation and generated queries. Since we utilize generative LLMs to do the extraction, traditional n-gram overlap-based generation metrics, such as BLEU and ROUGE, may not be a good indicator. Here we measure recall of annotation with a self-defined overlap between the reformulated query and annotation. In addition, to evaluate whether the reformulated query highlights the salient content, we calculate a self-defined metric called info ratio, and denote it as InfoR.

\subsubsection{Metric definition}
Given a query $Q$ and its reformulated format $U$ and annotation $A$, we first divide them into proper linguistic units. For example, separate keyword queries into words and separate original query with punctuation. Reformulated query $U$ is separated into $n$ units as $\{ u_1,u_2,...,u_n\}$ and the original query $q$ is separated into $m$ units  $\{ q_1,q_2,...,q_m\}$. We denote the annotation units $A_j$ within $q_j$  as $\{ a_{j1},a_{j2},...,a_{jk_j}\}$. Then we calculate Levenshtein distance to match each $q_j$  given  $u_i$ and calculate the character overlap between them. For each $u_i$, the recall of matched $A_j$, i.e. the overlap are calculated as follows:
\begin{gather}
    Overlap(u_i,A_j) = \sum_{p=1}^{j_k}\frac{|u_i \cap a_{jp}|}{|a_{jp}|},  \notag \\
    j=\arg\min\limits_{k} Levenshtein(u_i,q_k).
\end{gather}
Thus the overlap between $U$ and annotation $A$ are calculated as follows:
\begin{gather}
Overlap(U,A)=\sum_{i=1}^{n}Overlap(u_i,A_j), \notag \\
    j=\arg\min\limits_{k} Levenshtein(u_i,q_k).
\end{gather}
 The InfoR is defined to measure the ratio of salient content in reformulated query compared to original query. It is calculated as follows:
 \begin{equation}
InfoR(Q,U,A)=Overlap(U,A) \cdot \frac{|Q|}{|A|},
 \end{equation}
where $|A|$ and $|Q|$ denotes the length of annotation $A$ and original query $Q$, $U$ is the reformulated query.  $InfoR>1$ indicates the compression rate of salient content is lower than that of original query and the valuable information is more condensed. 

\begin{table}[t]
  \centering
  \caption{Statistic of reformulated queries}
    \begin{tabular}{llll}
    \toprule
    Query type & Avg. overlap & Avg. length & Avg. InfoR \\
    \midrule
    keyword & 34.46\% & 66.81 & 1.42 \\
    key sentence & 52.26\% & 219.45 & 1.07 \\
    summary & 43.99\% & 120.92 & 1.41 \\
    \bottomrule
    \end{tabular}%
  \label{tab:annot_recall}%
\end{table}%

\subsubsection{Analysis}
The statistics of the reformulated query are shown in Table~\ref{tab:annot_recall} . Combined with statistics in Table ~\ref{tab:main_results} and Table ~\ref{tab:annot_recall}, we observe that:(1) The $InfoR$ of all queries are greater than 1, indicating query reformulation methods above indeed highlight the salient content. Therefore, the model is more likely to capture valuable textual information for retrieval task and boost their performance. (2) A moderate overlap and length of a query may also important. Despite the $InfoR$ of key sentence is around 1, the retrieval models also improve on this kind of query, possibly due to the query length are much shorter than original while reserve moderate salient content to satisfy relevance judgment ~\cite{passagegain} . (3)  Query type influences the retrieval model. Since the keyword and summary type have a similar $InfoR$ and enough salient content to judge relevance, the summary type are generally better for retrieval models. The phenomenon implies the importance of fluent and complete format of query. (4) Query type command of prompts for LLMs also influence the generation quality of LLMS as both keyword and summary type have higher InfoR while key sentence type has a low InfoR. Although LLMs are powerful generative models, they may not be good at extraction tasks and satisfy all the commands in user prompts.

\section{Conclusion}
In this work, we investigate the potential of reformulating long queries to enhance legal case retrieval. We first analyze how traditional sparse and dense retrieval models attend to salient content in queries. We find that they attend to different content within a query and neither model's attention is consistent with human annotations. Therefore, we utilize LLMs to reformulate the legal case query by extracting and summarizing salient content from it. Experimental results show that the reformulated query improves the retrieval performance compared to the original query. Our findings and experiments provide new insight for legal case retrieval. 

There are also several potential limitations in this work. As legal search datasets are manually annotated, their scale is generally limited. We mainly analyse the models attention on a single dataset. Besides, we mainly focus on the attention characteristics of basic retrieval model and perform query content selection with LLMs to verify the potential to highlight valuable content. In the future work, we consider to expand the analysis to more advanced retrieval models on more legal datasets and we believe that with correct-designed methods, content selection methods is suitable to other long documents retrieval tasks. 

\begin{acks}
This work was supported by the National Natural Science Foundation of China (No. U21B2009 and 62302040) and the China Postdoctoral Science Foundation (No. 2022TQ0033).
\end{acks}

\bibliographystyle{ACM-Reference-Format}
\balance
\bibliography{sample-base}

\end{document}